\documentclass[conference]{IEEEtran}
\usepackage{cite}

\ifCLASSINFOpdf
\else
\fi
\usepackage[cmex10]{amsmath}
\usepackage{amsthm}
\usepackage{amssymb}
\usepackage{bbm}

\usepackage{array}
\usepackage{url}

\usepackage{pgfplots}
\usepackage{tikz}
\tikzstyle{block}=[draw=black, rectangle, fill=yellow!20, minimum size=2em,
rounded corners]
\tikzstyle{branch}=[fill,shape=circle, minimum size=3pt, inner sep=0pt]

\usepackage{enumerate}

\usepackage{flushend} 

\newtheorem{theorem}{Theorem}

\newtheorem{corollary}[theorem]{Corollary}
\newtheorem{lemma}[theorem]{Lemma}

\newtheorem{proposition}[theorem]{Proposition}

\theoremstyle{definition}

\newtheorem{definition}{Definition}

\theoremstyle{remark}
\newtheorem*{remark}{Remark}

\DeclareSymbolFont{symbolsC}{U}{ntxsyc}{m}{n}
\SetSymbolFont{symbolsC}{bold}{U}{ntxsyc}{b}{n}
\DeclareMathSymbol{\multimapinv}{\mathrel}{symbolsC}{18}

\newcommand{\Csiszar}{Csisz\'ar}

\newcommand{\Korner}{K\"orner}

\newcommand{\EE}{\mathbb{E}}
\newcommand{\NN}{\mathbb{N}}

\newcommand{\RR}{\mathbb{R}}
\newcommand{\RRR}{\bar{\mathbb{R}}}

\newcommand{\markov}{%
 \textnormal{\mbox{$\multimap\hspace{-0.73ex}-\hspace{-2ex}-$}}%
}

\newcommand{\dotle}{\mathrel{\dot{\le}}}

\newcommand{\IndexSet}[1]{[\![#1]\!]}

\DeclareMathOperator{\var}{var}
\DeclareMathOperator{\cov}{cov}

\newcommand{\KLDiv}{D}
\DeclareMathOperator*{\supp}{supp}

\newcommand{\abs}[1]{ {\left|#1\right|} }

\newcommand{\bX}{\mathbf{X}}
\newcommand{\bY}{\mathbf{Y}}
\newcommand{\bZ}{\mathbf{Z}}

\newcommand{\bx}{\mathbf{x}} 
\newcommand{\by}{\mathbf{y}} 
\newcommand{\bz}{\mathbf{z}}

\newcommand{\sfQ}{\mathsf{Q}}
\newcommand{\sfV}{\mathsf{V}}
\newcommand{\sfW}{\mathsf{W}}

\newcommand{\calA}{\mathcal{A}}

\newcommand{\calC}{\mathcal{C}}

\newcommand{\calP}{\mathcal{P}}

\newcommand{\calS}{\mathcal{S}}
\newcommand{\calT}{\mathcal{T}}

\newcommand{\calW}{\mathcal{W}}
\newcommand{\calX}{\mathcal{X}}
\newcommand{\calY}{\mathcal{Y}}
\newcommand{\calZ}{\mathcal{Z}}

\newcommand{\barE}{\bar{E}}

\newcommand{\bart}{\bar{t}}

\newcommand{\type}[1]{\hat{P}_{#1}}
\newcommand{\condtype}[1]{\hat{\sfV}_{#1}}

\begin{document}
%
\title{On the Secrecy Exponent of the Wire-tap Channel}

\author{%
  \IEEEauthorblockN{Mani~Bastani~Parizi and Emre~Telatar}
  \IEEEauthorblockA{%
    Information Theory Laboratory (LTHI),%
    EPFL, Lausanne, Switzerland \\%
    \{mani.bastaniparizi,emre.telatar\}@epfl.ch}%
}


%


\maketitle

\newcommand{\abstractcite}[1]{\IfFileExists{\jobname.bbl}{\cite{#1}}{*}}

\begin{abstract} 
  We derive an exponentially decaying upper-bound on the unnormalized amount of
  information leaked to the wire-tapper in Wyner's wire-tap channel setting. We
  characterize the exponent of the bound as a function of the  randomness
  used by the encoder. This exponent matches that of the recent work of Hayashi
  \abstractcite{hayashi2011} which is, to the best of our knowledge, the best
  exponent that exists in the literature. Our proof (like those of
  \abstractcite{han2014,hayashi2015}) is exclusively based on an i.i.d.\ random
  coding construction while that of \abstractcite{hayashi2011}, in addition,
  requires the use of random universal hash functions. 
\end{abstract}


%
\IEEEpeerreviewmaketitle

\section{Introduction}
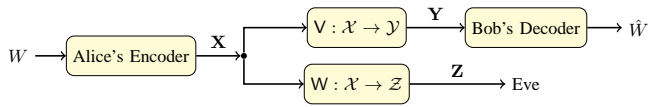
\begin{figure}[htb]
  \centering
  \scalebox{0.75}{
\begin{tikzpicture}[yscale=0.5, font=\small]
  \node		at 	(0,0)	(u)	{$W$};
  \node[block] 	at	(2,0) 	(enc)	{Alice's Encoder};
  \node[branch] at	(4,0) 	(dot)	{};
  \node[block]	at	(6,1)	(V)	{$\sfV: \calX \to \calY$};
  \node[block]	at	(6,-1)	(W)	{$\sfW: \calX \to \calZ$};
  \node[block]	at	(9,1)	(bobDec){Bob's Decoder};
  \node	at	(9,-1)	(eveDec){Eve};
  \node		at	(11,1)	(uhat) 	{$\hat{W}$};

  \draw[->,thick]	(u) -- (enc);
  \draw[->,thick]	(enc) -- node[above] {$\bX$} (dot);
  \draw[->,thick]	(dot) |- (V);
  \draw[->,thick]	(dot) |- (W);
  \draw[->,thick]	(V) -- node[above] {$\bY$} (bobDec);
  \draw[->,thick]	(W) -- node[above] {$\bZ$} (eveDec);
  \draw[->,thick]	(bobDec) -- (uhat);
\end{tikzpicture}}
  \caption{The Wire-Tap Channel}
  \label{fig:wiretapChannel}
\end{figure}

Wyner \cite{wyner1975} introduced the notion of the wire-tap channel
(Fig.~\ref{fig:wiretapChannel}) in 1975:
Alice wants to communicate a message $W \in \{1,\dots,M\}$ to Bob through a
communication channel $\sfV: \calX \rightarrow \calY$. Eve also has access to
what Alice transmits via a \emph{wire-tapper}'s channel $\sfW: \calX \rightarrow
\calZ$ and the aim of Alice is to keep the message hidden from her while
maximizing the rate of information transmitted to Bob, $R \triangleq \frac1n
\log M$.  

To this end, Alice encodes $W$ as a codeword $\bX \in \calX^n$ and sends it via
$n$ consecutive uses of the channel. Bob observes the output sequence of $\sfV$,
$\bY \in \calY^n$, and estimates $W$ given $\bY$.  On the other
side, Eve has access to $\bZ \in \calZ^n$ (the output sequence of $\sfW$), and
attempts to make an inference about $W$.

Wyner (in case when $\sfW$ is degraded with respect to $\sfV$) \cite{wyner1975}
and later \Csiszar{} and \Korner{} (in a more general context of $\sfV$ being
more capable than $\sfW$)\cite{csiszar1978} showed that, given any input
distribution $P_X$, Alice can communicate reliably to Bob at any rate $R$ up to
\begin{equation}
  I(X;Y) - I(X;Z), \label{eq:highestRate}
\end{equation}
(when $(X,Y) \sim P_X(x) \sfV(y|x)$ and $(X,Z) \sim P_X(x) \sfW(z|x)$) while
keeping the rate of information leaked to Eve about $W$ as small as desired;
i.e., guaranteeing
\begin{equation}
  \frac1n I(W;\bZ) \le \epsilon, \label{eq:weakSecrecy}
\end{equation}
for any $\epsilon > 0$, using sufficiently large $n$.

Wyner's measure of secrecy allows one to investigate the trade-off between
the message rate and the information leakage rate but is too weak from the
security point of view; even if the amount of information Eve learns about the
message $W$ normalized to the number of channel uses vanishes asymptotically,
the amount itself can grow unboundedly as the block-length increases. Therefore,
it is natural to remove the normalization factor in \eqref{eq:weakSecrecy} and
ask for \emph{strong secrecy}:
\begin{equation}
  I(W; \bZ) \le \epsilon.
\end{equation}
Maurer and Wolf showed that the highest achievable rate \eqref{eq:highestRate}
under \emph{strong secrecy} requirement does not change \cite{maurer2000}.

Classical achievability constructions \cite{wyner1975,massey1983} are based on
associating each message $w \in \{1,\dots,M\}$ with a sub-code of size $M' =
\exp(nR')$ and transmitting a randomly chosen codeword from that sub-code to
communicate $w$.  The reliability of the code is ensured by keeping the total
rate $R' + R$ below  $I(X;Y)$.  Furthermore, by varying the rate $R'$
from $0$ to $I(X;Z)$, the upper-bound on the information leakage rate,
$\frac1nI(W;\bZ)$, is controlled.  Particularly, by choosing the rate $R'$ just
\emph{below} $I(X;Z)$, weak secrecy is established. 

An alternative way to approach the secrecy problem is to establish secrecy
through \emph{channel resolvability} \cite{bloch2013,hayashi2006,hou2014}. 
Given an input distribution $P_X$ that induces the distribution $P_Z$ at the
output of a channel $\sfW: \calX \to \calZ$, a code of rate $I(X;Z)$ or larger
chosen from the i.i.d.\ $P_X$ random coding ensemble will, with high
probability, induce an output distribution that approximates $P_Z^n$ when
the index of the transmitted codeword is chosen uniformly at random. 
\cite{hayashi2006,hou2013,wyner1975CI,han1993,pierrot2013}.

For any fixed message $w \in \{1,\dots,M\}$ the output of Eve's channel has
distribution $P_{\bZ|W=w}$. It is not difficult to see that the secrecy is
guaranteed if $P_{\bZ|W=w}$ `well approximates' the product distribution
$P_Z^n$ by setting the sub-codes' rate $R'$ just \emph{above}
$I(X;Z)$. In particular, if we measure the quality of
approximation by asking the unnormalized Kullback-Leibler divergence between
$P_{\bZ|W=w}$ and $P_Z^n$ to be small, \emph{strong secrecy} will be
established.  Indeed, in \cite{hayashi2006,hou2014} it has been
shown that the information leakage, $I(W;\bZ)$ will be exponentially small in
$n$ provided that $R'$ is above $I(X;Z)$.  

\begin{definition} 
  Given $R$, $R'$ and $\sfW$, a number $E$
  is a \emph{secrecy exponent} for the wire-tapper channel
  $\sfW$, if there exist a sequence of reliable coding schemes of rate
  $R$, requiring the entropy rate $R'$ at the encoder, for which
  $\displaystyle\liminf_{n\to\infty}
  \textstyle -\frac{1}{n} \log[I(W;\bZ)] \ge E$.
\end{definition}

In \cite{hayashi2006,hou2014} the secrecy exponent is derived using i.i.d.\
random coding ensemble. More specifically, each message $w
\in \{1,\dots,M\}$ is associated with a sub-code whose codewords are
independently (and independent of the codewords of the other sub-codes) sampled
from the i.i.d.\ random coding ensemble.  The exponent is derived by
upper-bounding the ensemble-expectation of $\KLDiv(P_{\bZ|W} \Vert P_Z^n | P_W)$
and then concluding that there exists a sequence of codes in the ensemble using
which the information leakage decays at least as fast as $\EE[\KLDiv(P_{\bZ|W}
\Vert P_Z^n | P_W)]$ does.  The secrecy exponent of Hou and Kramer in
\cite{hou2014}  is derived based on their resolvability proof of
\cite[Section~III-A]{hou2013} which is simple but results in a small exponent.
However, by applying the method described in \cite[Section~III-B]{hou2013} to
the wire-tap channel setting a larger exponent can be obtained which is equal to
that of Hayashi in \cite{hayashi2006}. 

In \cite{hayashi2011}, Hayashi uses \emph{privacy amplification} to improve the
secrecy exponent based on a different construction than those of
\cite{hayashi2006,hou2013,hou2014}.  In addition to a code of size $M M'$, whose
codewords are sampled independently from the i.i.d.\ random coding ensemble, a
hash function is sampled from the ensemble of universal hash functions from
$\{1,\dots,M M'\}$ to $\{1,\dots,M\}$ and revealed to Alice, Bob, and Eve. A
message $m \in \{1,\dots,M\}$ is communicated by sending a randomly chosen
codeword from the code and, then, mapping the index of the sent codeword,  using
the hash function, to an element of $\{1,\dots,M\}$. The expected information
leakage (where the expectation is taken over both i.i.d.\ random coding
\emph{and} universal hash functions ensembles) is then upper-bounded to show
that the exponent of the bound is a secrecy exponent.

In this paper, we derive an exponentially decaying upper-bound on
$\EE[\KLDiv(P_{\bZ|W=w} \Vert P_Z^n)]$, where the expectation is taken over the
i.i.d.\ random coding ensemble (i.e., the construction used in
\cite{hayashi2006,hou2013,hou2014}), by analyzing the deviations of
$P_{\bZ|W=w}$ from its mean.  It then follows (by standard expurgation
arguments) that for $\forall \epsilon > 0$,  there exist a  code of essentially
the same rate $R$, using which $\max_{w} \KLDiv(P_{\bZ|W=w} \Vert P_Z^n) \le
(1+\epsilon) \EE[\KLDiv(P_{\bZ|W=w} \Vert P_Z^n)]$. As already noted in
\cite{hou2014}, this is a \emph{worst-case} measure of secrecy in contrast to
$I(W;\bZ)$ which is an
average-case measure of secrecy. In addition, this shows that our lower-bound on
$\lim_{n \to \infty} -\frac1n \log \EE[\KLDiv(P_{\bZ|W=w} \Vert P_Z^n)]$ is a
secrecy exponent. This exponent matches that of \cite{hayashi2011} which is
larger than those of \cite{hayashi2006,hou2013,hou2014}. 

\section{Notation}
We use uppercase letters (like $X$) to denote a
random variable and corresponding lowercase version ($x$) for a realization of
that random variable. The boldface letters denote sequences of length $n$. The
$i$-th element of a sequence $\bx$ is denoted as $x_i$.  
We denote finite sets by script-style uppercase letters like $\mathcal{S}$. The
cardinality of set $\calS$ is denoted by $\abs{\calS}$. For a positive integer
$m$, $ \IndexSet{m} \triangleq \{1,2,\dots,m\}$.  $\RR$ denotes the set of real
numbers and $\RRR = \RR \cup \{-\infty, +\infty\}$ is the set of \emph{extended}
real numbers.  We write $f(n) \doteq g(n)$ (resp.\ $f(n)  \dotle g(n)$) if
$\lim_{n \to \infty} \frac1n \log \frac{f(n)}{g(n)} = 0$ (resp.\ $\le 0$).

We denote the set of distributions on alphabet $\calX$ as $\calP(\calX)$. 
If $P \in \calP(\calX)$,  $P^n \in \calP(\calX^n)$ denotes the
product distribution  $P^n(\bx) \triangleq \prod_{i=1}^n P(x_i)$.
Likewise, if $\sfV: \calX \to \calY$ is a conditional distribution
$\sfV^n:\calX^n \to \calY^n$ denotes the conditional distribution 
$\sfV^n(\by|\bx) = \prod_{i=1}^n \sfV(y_i|x_i)$. 

We denote the \emph{type} of a sequence $\bx \in \calX^n$ by $\type{\bx} \in
\calP(\calX)$ and the \emph{conditional type} of $\by \in \calY^n$ given $\bx
\in \calX^n$ by $\condtype{\by|\bx}: \calX \to \calY$ (see
\cite[Chapter~2]{csiszar2011IT} for formal definitions).  

A distribution $\hat{P} \in \calP(\calX)$ is an \emph{$n$-type} if $n
\hat{P}(x) \in \NN_{\ge 0}$ for $\forall x \in \calX$.
We denote the set of $n$-types on $\calX$ as $\hat{\calP_n}(\calX) \subsetneq
\calP(\calX)$ and use the fact that $|\hat\calP_n(\calX)| =
O(n^{\abs{\calX}})$ \cite[Lemma~2.2]{csiszar2011IT} repeatedly.

If $\hat P \in \hat{\calP}_n(\calX)$, we denote the set of all sequences of type
$\hat P$ as $\calT_{\hat{P}} \subset \calX^n$. If ${\hat \sfV}: \calX \to \calY$
is a conditional distribution, the \emph{${\hat \sfV}$-shell} of $\bx \in
\calX^n$, is denoted as $\calT_{\hat \sfV}(\bx) \subset \calY^n$.
\section{Result}
In the rest of the paper $(X,Z) \in \calX \times \calZ$ denotes the pair of
random variables whose joint distribution is $P_{X,Z}(x,z) = P_X(x) \sfW(z|x)$
where $P_X$ is a fixed input distribution.  For simplicity (and with no
essential loss of generality) we assume the $\supp(P_X) = \calX$ and
$\supp(P_Z) = \calZ$.\footnote{The second assumption follows from the first
  together with the assumption that for $\forall z \in \calZ$ there exist at
least one $x$ such that $\sfW(z|x) > 0$.}

Following \cite{massey1983} we consider the following random code construction:
for every message $w \in \IndexSet{M}$, a codebook of size $M' \triangleq \exp(n
R')$, denoted by $\calC_w$, is constructed by sampling $M'$ codewords,
$\bX_{w,w'}, w' \in \IndexSet{M'}$ independently from the product distribution
$P_X^n$. In order to communicate the message $w$, Alice picks $w' \in
\IndexSet{M'}$ uniformly at random and transmits $\bX_{w,w'}$.  Given such a
construction, for every $w \in \IndexSet{M}$ and  $\bz \in \calZ^n$,  the
conditional output distribution of $\sfW$ is
\begin{equation}
  P_{\bZ|W}(\bz|w)= \frac{1}{M'} \sum_{w'=1}^{M'}
  \sfW^n\big(\bz|\bX_{w,w'}\bigr),
  \label{eq:condProbZ}
\end{equation}
which is an average of i.i.d.\ random variables and
\begin{equation} 
  \EE \bigl[P_{\bZ|W}(\bz|w)\bigr] =  P_Z^n(\bz), \qquad \forall w \in
  \IndexSet{M}.
  \label{eq:expectedPZW}
\end{equation}
\begin{theorem} \label{thm:main}
  Using the aforementioned construction, for $\forall w \in \IndexSet{M}$,
  \begin{equation*}
    \EE \bigl[ \KLDiv(P_{\bZ|W = w} \Vert P_Z^n) \bigr] 
    \dotle \exp[-n E_{\rm s}(P_X,\sfW,R')].
  \end{equation*}
  with 
  \begin{equation}
    E_{\rm s}(P_X,\sfW, R') = \max_{0 \le \lambda \le 1} \bigl\{
    \lambda R' - F_0(P_X,\sfW,\lambda) \bigr\},
    \label{eq:generalEs}
  \end{equation}
  where
  \begin{equation*}
    F_0(P_X,\sfW,\lambda)
    \triangleq
    \log \biggl[ \sum_{z \in \calZ} P_Z(z)
    \sum_{x \in \calX} P_{X|Z}(x|z)^{1+\lambda} P_X(x)^{-\lambda} \biggr].
  \end{equation*}
\end{theorem}
\begin{remark} $F_0(P_X,\sfW,\lambda)$ is a convex function of $\lambda$ (cf.
  Appendix~\ref{app:f0convex}) passing through the origin with the slope 
  \begin{equation*}
    \frac{\partial}{\partial \lambda} F_0(P_X,\sfW,\lambda) \Big|_{\lambda = 0}
    = I(X;Z).
  \end{equation*}
  Hence $E_{\rm s}(P_X,\sfW,R') \ge 0$ with equality iff $R' \le I(X;Z)$.
\end{remark}

The only random quantity involved in the divergence $\KLDiv(P_{\bZ|W = w} \Vert
P_Z^n)$ is the conditional distribution $P_{\bZ|W=w}$ whose expectation is
$P_Z^n$ as shown in \eqref{eq:expectedPZW}. To prove Theorem~\ref{thm:main} we
shall analyze the deviations of the random variables $P_{\bZ|W}(\bz|w)$ from
their mean, $P_Z^n(\bz)$.

As an immediate corollary to Theorem~\ref{thm:main} we have:
\begin{corollary} \label{cor:existence}
  For any input distribution $P_X$ and a pair of rates $R$ and $R'$,
  there exists a reliable code of rate $R$ using which, for any message
  distribution $P_W$,
  \begin{align*}
    P_{\rm e} &\dotle \exp[-n E_{\rm r}(P_X, \sfV, R+R')], \\
    I(W;\bZ) &\dotle \exp[-n E_{\rm s}(P_X, \sfW,R')],
  \end{align*} 
  where $P_{\rm e}$ denotes the decoding error probability of Bob and $E_{\rm
  r}$ is Gallager's random coding exponent \cite[Chapter~5]{gallager1968}.
  Hence, for $(R,R')$ such that $R+R' < I(X;Y)$, the $E_{\rm s}$ in
  Theorem~\ref{thm:main} is a secrecy exponent.  
\end{corollary}
Corollary~\ref{cor:existence} is proved in Appendix~\ref{app:existence}.
\section{Proof of Theorem~\ref{thm:main}} \label{sec:proof}
For $\forall w \in \IndexSet{M}$ and $\forall \bz \in \calZ^n$ let
\begin{equation}
  U_n(\bz|w) \triangleq \frac{P_{\bZ|W}(\bz|w)}{P_Z^n(\bz)}.
  \label{eq:udef}
\end{equation} 
Using \eqref{eq:expectedPZW}, it is easy to see that $\EE[U_n(\bz|w)] = 1$.

Using the linearity of expectation, we have:
\begin{align}
  & \EE\bigl[\KLDiv(P_{\bZ|W=w} \Vert P_Z^n)\bigr] \nonumber \\
  & \quad = \sum_{\bz \in \calZ^n} \EE\Bigl[P_{\bZ|W}(\bz|w) \log \Bigl(
    \frac{P_{\bZ|W}(\bz|w)}{P_Z^n(\bz)} \Bigr)
  \Bigr]
  \nonumber \\
  & \quad = \sum_{\bz \in \calZ^n} P_Z^{n}(\bz) \EE\bigl[ U_n(\bz|w)
  \log \bigl( U_n(\bz|w) \bigr) \bigr] \nonumber  \\
  & \quad = \sum_{\hat P \in \hat\calP_n(\calZ)} 
  \sum_{\bz \in \calT_{\hat P}} P_Z^n(\bz) \EE\bigl[U_n(\bz|w) \log \bigl(
  U_n(\bz|w) \bigr) \bigr].
  \label{eq:typeSum}
\end{align}

To prove Theorem~\ref{thm:main}, we shall use the following result.
\begin{lemma} \label{lem:typeBound}
  For $P \in \calP(\calZ)$, let
  \begin{align}
    & G_0(P_{X,Z}, P, \lambda) \nonumber \\
    & \quad \triangleq \sum_{z \in \calZ} P(z) \log\Bigl[\sum_{x\in\calX}
    P_{X|Z}(x|z)^{1+\lambda} P_X(x)^{-\lambda} \Bigr], \label{eq:g0def}
  \end{align}
  and
  \begin{equation}
    E_t(P_{X,Z}, R', P) \triangleq \max_{0 \le \lambda \le 1} \bigl\{ \lambda R'
    - G_0(P_{X,Z}, P, \lambda)\bigr\}. \label{eq:EtVal}
  \end{equation}
  Then, for every $w \in \IndexSet{M}$,
  \begin{align}
    & \EE\bigl[U_n(\bz|w) \log \bigl(U_n(\bz | w) \bigr) \bigr] \nonumber \\
    & \qquad \dotle 
    \exp[-n E_t(P_{X,Z}, R', \type{\bz})].  \label{eq:typeBound}
  \end{align}
\end{lemma}
Having proved Lemma~\ref{lem:typeBound}, Theorem~\ref{thm:main} follows by using
\eqref{eq:typeBound} in \eqref{eq:typeSum} and \cite[Lemma~2.6]{csiszar2011IT}
to conclude 
\begin{equation*}
  \EE\bigl[ \KLDiv(P_{\bZ|W=w} \Vert P_Z^n) \bigr] \dotle 
  \exp\bigl[ -n E_{\rm s}(P_X,\sfW,R') \bigr],
\end{equation*}
where
\begin{align} 
  & E_{\rm s}(P_X, \sfW, R') \nonumber \\
  & \quad \triangleq \min_{P \in \calP(\calZ)} \{ \KLDiv(P
    \Vert P_Z) + E_t(P_{X,Z} ,R', P) \}.
  \label{eq:exponentDef}
\end{align}
Using \eqref{eq:EtVal}, the equivalence of \eqref{eq:exponentDef} and
\eqref{eq:generalEs} is shown in Appendix~\ref{app:secExp}. This completes the
proof of Theorem~\ref{thm:main}. \hfill \IEEEQED

\begin{IEEEproof}[Proof of Lemma~\ref{lem:typeBound}]
  Pick any $\hat P \in \hat\calP_n(\calZ)$ and observe that for $\bz \in
  \calT_{\hat{P}}$,
  \begin{equation*}
    \frac{\sfW^n(\bz|\bx)}{P_Z^n(\bz)} 
    = 
    \exp\bigl[n \bigl(\KLDiv(\condtype{\bx | \bz} \Vert P_X | \hat P) -
    \KLDiv(\condtype{\bx | \bz} \Vert P_{X|Z} | \hat P)\bigr)].
  \end{equation*}
  For every $P \in \calP(\calZ)$ and stochastic matrix $\sfQ:
  \calZ \to \calX$ define
  \begin{equation}
    A_{X,Z}(P;\sfQ)  \triangleq  
    \KLDiv(\sfQ \Vert P_X | P) - \KLDiv(\sfQ \Vert P_{X|Z} | P).
    \label{eq:apqdef} 
  \end{equation}
  Thus, using \eqref{eq:condProbZ},
  \begin{equation}
    U_n(\bz|w) = \frac{1}{M'} \sum_{w'=1}^{M'} \exp\bigl[n A_{X,Z}(\hat P;
    \condtype{\bX_{w,w'} | \bz})  \bigr] \label{eq:usum}
  \end{equation}
  Let
  \begin{equation}
    \tilde\calA \triangleq \big\{%
      A_{X,Z}(\hat{P}; \hat\sfQ) \text{ for all conditional types $\hat\sfQ$}
    \bigr\} \subset \RRR, \label{eq:atildedef}
  \end{equation}
  and observe that $|\tilde\calA| =  O(n^{\abs{\calX} \abs{\calZ}})$.
  Set $\calA \triangleq \{a \in \tilde\calA: a > -\infty\}$
  and for each $a \in \calA$ define
  \begin{equation}
    \calT_{a}(\bz) \triangleq \bigcup_{\hat\sfQ: A_{X,Z}(\hat{P};\sfQ) = a} 
    \calT_{\hat\sfQ}(\bz) \subseteq \calX^n,
    \label{eq:tdef}
  \end{equation} 
  where $\calT_{\hat\sfQ}(\bz)$ is the $\hat\sfQ$-shell of $\bz$ and the union
  is over conditional types $\hat\sfQ : \calZ \to \calX$ (thus contains
  $O(n^{\abs{\calX}\abs{\calZ}})$ shells).
  Now we can  rewrite \eqref{eq:usum} as\footnote{Since $\bz$ and $w$ are
    assumed to be fixed throughout the proof, we drop them from the argument of
  $U_n$ for the sake of brevity.}
  \begin{equation}
    U_n \triangleq 
    U_n(\bz|w) = \frac1{M'} \sum_{a \in \calA} N_a \exp(n a),
    \label{eq:usuma}
  \end{equation}
  with $N_a \triangleq \abs{\left\{w': \bX_{w,w'} \in \calT_a(\bz)\right\}}$
  denotes the number of codewords of $\calC_w$ in $\calT_a(\bz)$. Since
  the codewords are independent, $N_a$ is a $\mathrm{Binomial}(M',p_a)$ random
  variable where,
  \begin{align}
    p_a & = P_X^n\bigl(\calT_a(\bz)\bigr)
    = \sum_{\hat\sfQ: A_{X,Z} (\hat{P};\hat\sfQ) = a} P_X^n\bigl(
    \calT_{\hat\sfQ}(\bz)\bigr) \nonumber \\
    & \doteq \exp\Bigl[ - n  \min_{\hat\sfQ: A_{X,Z}(\hat{P}; \hat\sfQ) =
    a} \KLDiv(\hat\sfQ \Vert P_X | \hat{P})   \Bigr]. \label{eq:pa} 
  \end{align}
  In the above,  the second equality follows since $\hat\sfQ$-shells are
  disjoint, the third equality follows from \cite[Lemma 2.6]{csiszar2011IT} (a
  similar approach is used in \cite{merhav2014} to express a quantity of
  interest as a weighted sum of Binomial random variables).

  In Appendix~\ref{app:eb} we compute the value of  
  \begin{equation}
    E_b(P_{X,Z}, P, a) \triangleq \min_{\hat\sfQ: A_{X,Z}(P;\hat\sfQ) = a} 
    \KLDiv(\hat\sfQ \Vert P_X | P) \label{eq:EbDef}
  \end{equation}
  and, in particular, show that 
  \begin{equation}
    E_b(P_{X,Z}, P, a) \ge a, \label{eq:EbBound}
  \end{equation}
  with equality iff $a = \KLDiv(P_{X|Z} \Vert P_X | P)$.

  Partition $\calA = \calA_1 \cup \calA_2$ as
  \begin{equation*}
    \calA_1 = \{a \in \calA: a \le R'\}, \qquad
    \calA_2 = \{a \in \calA: a > R'\},
  \end{equation*}
  and split \eqref{eq:usuma} as 
  \begin{equation*}
    U_n  = \underbrace{%
      \frac{1}{M'} \sum_{a \in \calA_1} N_a \exp(na)
    }_{\triangleq S_n}
    +
    \underbrace{%
      \frac{1}{M'} \sum_{a \in \calA_2} N_a \exp(na)
    }_{\triangleq T_n}.
  \end{equation*} 

  For non-negative $s$ and $t$ and $u \triangleq  s+t$ we have
  \begin{align*}
    u \ln (u)  & = s \ln (u) + t \ln (u)  \\
    & = s \ln (s) + s \ln (1 + t/s) + t \ln (u) \\
    & \le s \ln (s) + t (1 + \ln(u))
  \end{align*}
  where the inequality follows since $\ln(1+t/s) \le t/s$. 
  Hence,
  \begin{align}
    & \EE[U_n \log(U_n) ] \doteq  \EE[U_n \ln(U_n)]
    \nonumber \\
    & \quad \le \EE[S_n
    \ln(S_n)]+ \EE\bigl[T_n \bigl(1 + \ln(U_n) \bigr)\bigr].
    \label{eq:ebound1}
  \end{align}
  Moreover, since $U_n \le 1/ P_Z^n(\bz)$, we have 
  \begin{equation*}
    \ln(U_n) \le \ln \bigl(1/P_Z^n(\bz)\bigr) \le n \ln
    (1/p_0)
  \end{equation*}
  where $p_0 \triangleq \min_{z \in \calZ} P_Z(z) > 0$. Thus,  from
  \eqref{eq:ebound1} we have
  \begin{align}
    \EE\bigl[U_n \ln(U_n) \bigr] & \le  \EE[S_n \ln(S_n)]
     + (n \ln(1/p_0)  + 1 ) \EE[T_n]  \nonumber \\
     & \doteq \EE[S_n \ln(S_n)] + \EE[T_n].
    \label{eq:ebound2}
  \end{align}

  We now upper-bound each of the above expectations to complete the proof.

  First we note that for any constant $c \in \RR$,
  \begin{equation}
    \EE[S_n \ln (S_n)] = \EE\bigl[S_n \ln (S_n) + c(S_n - \EE[S_n])\bigr].
    \label{eq:tilte}
  \end{equation}
  In particular,
  \begin{equation*}
    \EE[S_n \ln(S_n)] = \EE[\psi(S_n)]
  \end{equation*}
  where
  \begin{equation}
    \psi(s) \triangleq s \ln(s) - \bigl( \ln \bigl( \EE[S_n] \bigl) + 1
    \bigr)(s-\EE[S_n]).
    \label{eq:psiDef}
  \end{equation}
  One can check that (see Fig.~\ref{fig:psi})
  \begin{equation}
    \psi(s) \le \frac{(s - \EE[S_n])^2}{\EE[S_n]} + \EE[S_n] \ln(\EE[S_n]) 
    \le \frac{(s - \EE[S_n])^2}{\EE[S_n]},
    \label{eq:psiBound}
  \end{equation}
  where the last inequality follows since $\EE[S_n] = 1 - \EE[T_n] \le 1$ as
  $S_n$ and $T_n$ are both non-negative random variables.
 
  \begin{figure}[t]
    \centering
    \begin{tikzpicture}[font=\small]
  \begin{axis}[%
      height = 0.2\textheight, width=\columnwidth,
      enlargelimits = false,
      ymin=-0.375,
      axis y line=center, axis x line=middle,
      xlabel = $s$,
      legend pos = north west,
      xtick={0,0.75}, xticklabels={$0$,{$\overline{S_n}$}}, 
      xticklabel style={yshift=0.5ex, anchor=south},
      yticklabel style={rotate=0, anchor = east},
      ytick={-0.21576,0}, yticklabels={$\overline{S_n} \ln(\overline{S_n})$},
      grid=none,
      domain=0:1.875,
    ]
    
    \addplot[thick,smooth] {x*ln(x) - (1+ln(0.75))*(x-0.75)};
    \addlegendentry{$\psi(s)$}; 
    \addplot[very thick, blue, dashed, smooth]
    {4/3*(x-0.75)^2+0.75*ln(0.75)};
    \addlegendentry{$(s-\overline{S_n})^2/\overline{S_n} + \overline{S_n} \ln
    (\overline{S_n})$}; 
    \addplot[very thick, green!50!black, dotted,smooth]
    {4/3*(x-0.75)^2};
    \addlegendentry{$(s-\overline{S_n})^2/\overline{S_n}$}; 
    \draw[gray,dashed] (axis cs: 0,-0.21576) -- (axis cs: 0.75,-0.21576);
    \draw[gray,dashed] (axis cs: 0.75,0) -- (axis cs: 0.75,-0.21576);
  \end{axis}
\end{tikzpicture}
    
    \caption{The function $\psi(s)$ defined in \eqref{eq:psiDef} and the
    upper-bound in \eqref{eq:psiBound}. In the figure $\overline{S_n} \triangleq
    \EE[S_n]$.}
    \label{fig:psi}
  \end{figure}

  Using \eqref{eq:psiBound} in \eqref{eq:tilte} we conclude that 
  \begin{equation}
    \EE[S_n \ln(S_n)] \le \frac{\var(S_n)}{\EE[S_n]}.
    \label{eq:ESlogSBound}
  \end{equation}

  We now have,
  \begin{align}
    \EE[S_n] & = \sum_{a \in \calA_1} p_a \exp(n a) \nonumber \\
    & \doteq \exp\Bigl[-n \min_{a \in \calA_1} \bigl\{E_b(P_{X,Z}, \hat P, a) -
    a \bigr\} \Bigr], \label{eq:ESBound}
  \end{align}
  where the last equality follows since $|\calA_1| =
  O(n^{\abs{\calX}\abs{\calZ}})$. Furthermore,
  \begin{align}
    & \var(S_n)  = \frac{1}{{M'}^2} \sum_{(a,a') \in \calA_1^2}
    \exp[n (a+a')] \cov(N_a, N_{a'}) \nonumber \\ 
    & \quad \stackrel{\text{(a)}}{\le} \frac{1}{{M'}^2} \sum_{(a,a') \in
    \calA_1^2} \exp[n(a+a')] \sqrt{\var(N_a)} \sqrt{\var(N_{a'})} \nonumber \\
    & \quad = \frac{1}{{M'}^2} \left(\sum_{a \in \calA_1}
    \exp[n a] \sqrt{\var(N_a)} \right)^2  \nonumber \\
    & \quad \stackrel{\text{(b)}}{\doteq} \frac{1}{{M'}^2} 
    \left(\max_{a \in \calA_1} \left\{ \exp[n a] \sqrt{\var(N_a)}\right\}
    \right)^2 \nonumber \\
    & \quad =  \max_{a \in \calA_1}\Bigl\{ \frac{1}{ {M'}^2 } \exp[2 n a]
    \var(N_a) \Bigr\} \nonumber \\
    & \quad \stackrel{\text{(c)}}{\le} 
    \max_{a \in \calA_1} \Bigl\{ \frac{1}{M'} \exp[2 n a] p_a \Bigr\}
    \nonumber \\
    & \quad \doteq \exp\Bigl[ -n \min_{a \in \calA_1} \bigl\{R' + E_b(P_{X,Z},
    \hat P, a) - 2 a \bigr\} \Bigr]. \label{eq:varSBound}
  \end{align}
  In the above, 
  \begin{enumerate}[(a)]
    \item follows by Cauchy--Schwarz inequality,
    \item follows since $|\calA_1| = O(n^{\abs{\calX}{\abs{\calZ}}})$,
    \item follows since $\var(N_a) = M' p_a (1-p_a) \le M' p_a$,
  \end{enumerate}
  and finally \eqref{eq:varSBound} follows from \eqref{eq:pa} and
  \eqref{eq:EbDef}.

  Similar to \eqref{eq:ESBound}, 
  \begin{equation}
    \EE[T_n] \doteq  \exp\Bigl[ - n  \min_{a \in \calA_2} \bigl\{
    E_b(P_{X,Z},\hat P, a) - a \bigr\} \Bigr]. \label{eq:ETBound}
  \end{equation}

  Putting \eqref{eq:ESBound} and \eqref{eq:varSBound} in \eqref{eq:ESlogSBound}
  together with \eqref{eq:ETBound} in \eqref{eq:ebound2} we conclude that
  \begin{align}
    E_t(P_{X,Z}, R', \hat P)  = &\min\{E_1(P_{X,Z}, R', \hat P) -
    \barE_2(P_{X,Z}, R', \hat P), \nonumber \\
    & \qquad E_2(P_{X,Z}, R', \hat P)\}, \label{eq:typeExpMin1}
  \end{align}
  where
  \begin{align}
    E_1(P_{X,Z}, R', \hat P)  & \triangleq \min_{a \le R'} \bigl\{
      R' + E_b(P_{X,Z}, \hat P, a) - 2 a \bigr\}, \label{eq:e1def} \\
    E_2(P_{X,Z}, R', \hat P)  & \triangleq \min_{a  > R'} \bigl\{
      E_b(P_{X,Z},  P, a) - a \bigr\}, \label{eq:e2def} \\
    \barE_2(P_{X,Z}, R', \hat P)  & \triangleq \min_{a  \le R'} \bigl\{
      E_b(P_{X,Z},  P, a) - a \bigr\}. \label{eq:e2bardef}
  \end{align}

  We now observe that:
  \begin{enumerate}[i.]
    \item lower-bounding $R'$ by $a$ in \eqref{eq:e1def} shows $E_1(P_{X,Z},
      R',\hat P) - \barE_2(P_{X,Z}, R', \hat P) \ge 0$.
    \item by \eqref{eq:EbBound}, one and only one of $E_2(P_{X,Z}, R', \hat P)$
      or $\barE_2(P_{X,Z}, R', \hat P)$ is zero.
  \end{enumerate}
  Thus \eqref{eq:typeExpMin1} simplifies to 
  \begin{equation}
    E_t(P_{X,Z}, R', \hat P) = 
    \min \bigl\{E_1(P_{X,Z}, R', \hat P), E_2(P_{X,Z}, R', \hat P)\bigr\}
    \label{eq:typeExpMin}
  \end{equation}

  In Appendix~\ref{app:e1e2} we show that
  \begin{subequations}
    \begin{align}
      E_1(P_{X,Z},R', \hat P) &= \max_{\lambda \le 1} \bigl\{ \lambda R' -
	G_0(P_{X,Z}, \hat P, \lambda) \bigr\}, \label{eq:e1val} \\
      E_2(P_{X,Z},R', \hat P) &= \max_{\lambda \ge 0} \bigl\{ \lambda R'-
	G_0(P_{X,Z}, \hat P, \lambda) \bigr\}. \label{eq:e2val}
    \end{align}
  \end{subequations}
  Using the above in \eqref{eq:typeExpMin} concludes the proof.
\end{IEEEproof}
\section{Discussion}
We derived a lower-bound on the secrecy exponent of the wire-tap
channel using i.i.d.\ random codes. Comparing \eqref{eq:generalEs} with
\cite[Equation~(12)]{hayashi2011}, we see that our exponent is equal to that of
\cite{hayashi2011} which is the best lower-bound on the secrecy exponent among
those reported in \cite{hayashi2006,hayashi2011,hou2014}.  However, our
proof is based on a pure i.i.d.\ random coding construction and does not require
the ensemble of universal hash functions as an additional tool. While this
manuscript was in review, it came to our attention that in
\cite{han2014,hayashi2015} also alternative derivations of the same lower-bound
are given based on pure i.i.d.\ random coding constructions.

Our proof is a generalization of that of \cite[Section~III-A]{hou2013}; instead
of partitioning the set of output sequences $\calZ^n$ into two classes of
typical and atypical sequences, we partition it into $O(n^{|\calZ|})$
type-classes to upper-bound the expected unnormalized Kullback-Leibler
divergence between the output distribution and the desired product distribution
$P_Z^n$.  In addition, in Lemma~\ref{lem:typeBound}, we bound the point-wise
difference between those distributions at each $\bz \in \calZ^n$. 

Furthermore, we believe that the method described here has merit in showing the
doubly exponential nature of the concentration of the output distribution; as we
see in \eqref{eq:condProbZ}, the output distribution $P_{\bZ|W}(\bz|w)$ is an
average of $M'$ i.i.d.\ random variables. If the distribution of the summands
was independent of $M'$, the average would have concentrated around its mean
exponentially fast in $M'$, that is \emph{doubly exponentially fast} in $n$.
Although this is not the case, we see in the proof of Lemma~\ref{lem:typeBound}
that among polynomially many summands in \eqref{eq:usuma}, only the one
corresponding to $a = \KLDiv(P_{X|Z} \Vert P_X | \type{\bz})$ has a
significant contribution to the mean of $U_n(\bz|w)$ (which is a normalized
version of $P_{\bZ|W}(\bz|w)$); the rest all have exponentially small means.
Applying the Chernoff bound to this particular term, 
we see that if $R' > \KLDiv(P_{X|Z}\Vert
P_X | \type{\bz})$ the dominant term concentrates around its mean doubly
exponentially fast in $n$. In particular, there exists a class of wire-tapper
channels for which $U_n(\bz|w)$ consists only of this dominant
term.\footnote{This happens if for $\forall z \in \calZ$, for every $x \in
  \calX$ either $\sfW(z|x) = 0$ or $\sfW(z|x) = \epsilon_z$ for some constant
$\epsilon_z < 1$ independent of $x$.}

The achievability constructions of
\cite{hayashi2006,hayashi2011,hou2013,hou2014,han2014,hayashi2015} are based on
i.i.d.\ random codes.  It is an open question whether random
constant-composition codes \cite{csiszar2011IT} will lead to a better secrecy
exponent.  We believe that our method is easily adaptable to other types of
random coding (some ideas presented in \cite{hayashi2011allerton} can also be
useful in this direction).  Another important subject in the context of wire-tap
channel is to derive non-trivial upper-bounds on the secrecy exponent.

The performance of a wire-tap code is measured via two quantities, the error
probability and the information leakage, which are both shown to be
exponentially decaying as a function of the block-length $n$.  The trade-off
between these exponents has been recently studied in \cite{Tan2015}.

We conclude our discussion by remarking that, as shown in \cite{csiszar1978},
for general channels $\sfV$ and $\sfW$, any message rate up to
\begin{equation*}
  I(V; Y) - I(V; Z),
  \end{equation*}
where $V \markov X \markov (Y,Z)$ form a Markov chain, is achievable. Our
results (and also those of others cited) are straightforwardly extensible to the
case when the channels are prefixed with a channel $P_{X|V}$ and auxiliary
random variable $V$ is used.

\section*{Acknowledgment}
The authors would like to thank Prof. Neri Merhav, Prof. Vincent Y. F. Tan,  and
Mohammad Hossein Yassaee for their helpful comments on an earlier version of
this work.

This work was supported by the Swiss NSF under grant number 200020\_146832.



\bibliographystyle{IEEEtran}
\bibliography{IEEEfull,%
  bibliography/books,%
  bibliography/journals,%
  bibliography/conferences%
}
%

\appendices
\section{Proof of \eqref{eq:expectedPZW}} \label{app:expectedP}
The right-hand-side of \eqref{eq:condProbZ} is the average of identically
distributed random variables.  The mean of each of them is
\begin{align*}
  & \EE \bigg[ \prod_{i=1}^{n} \sfW\left(z_i | X_i\right) \bigg]
  = \prod_{i=1}^{n} \EE_{X \sim P_X} \left[\sfW\left(z_i | X\right) \right] \\
  & \qquad = \prod_{i=1}^{n} \Big[ \sum_{x \in \calX} P_X(x) \sfW(z_i|x)
  \Big] = \prod_{i=1}^{n} P_Z(z_i)
\end{align*}
In the above, the first equality follows since the codewords are sampled from
the product distribution $P_X^n$. \hfill \IEEEQED
\section{Proof of Corollary~\ref{cor:existence}} \label{app:existence}
Let $M \triangleq \exp(nR)$ and construct $2M$ i.i.d.\ codebooks of size $M'
\triangleq \exp(n R')$, $\calC_w, w \in \IndexSet{2M}$ by sampling each codeword
independently from the product distribution $P_X^n$. As we already described, in
order to communicate $w \in \IndexSet{2M}$, Alice picks $w' \in \IndexSet{M'}$
uniformly at random and transmits $\bX_{w, w'}$ over the channel.  The union of
this codebooks $\calC \triangleq \bigcup_{w \in \IndexSet{2M}} \calC_w$ is a
random i.i.d.\  codebook of rate $R' + R + \frac{\log(2)}{n}$. Hence, using this
ensemble for communicating over $\sfV$, for each $w \in \IndexSet{2M}$, the
expected decoding error probability is upper-bounded as 
\begin{align} &
  \EE\bigl[ \Pr[\hat{W} \ne W | W = w] \bigr] \nonumber \\ & \quad \le \EE\bigl[
    \Pr[\{\hat{W} \ne W\} \cup \{\hat{W'} \ne W'\} | W = w \bigr] \nonumber \\ &
    \quad \le \exp\bigl[-n E_{\rm r}\bigl(P_X,\sfV, R+R'+ o(1)\bigr)\bigr],
\end{align}
due to \cite[Theorem~5.6.2]{gallager1968}.
In the above, $\hat{W}$ and $\hat{W'}$ denote, respectively, the maximum
likelihood estimations of $W$ and $W'$ given $\bY$, the output sequence of
$\sfV$. Consequently,
\begin{align}
  & \EE\Bigl[ \frac{1}{2M} \sum_{w=1}^{2M} \Pr[\hat{W} \ne w | W = w] \Bigr]
  \nonumber \\ & \qquad \dotle \exp[-n E_{\rm r}(P_X,\sfV, R+R')].
\end{align}
Likewise, Theorem~\ref{thm:main} implies 
\begin{align}
  & \EE \Bigl[ \frac1 {2M} \sum_{w=1}^{2M} \KLDiv(P_{\bZ|W=w} \Vert
  P_Z^n) \Bigr]  \nonumber \\
  & \qquad \dotle \exp[-n E_{\rm s}(P_X,\sfW,R')].
\end{align}
Therefore, there exists a code $\calC^* = \bigcup_{w \in \IndexSet{2M}}
\calC_w^*$ in the ensemble  using which we simultaneously
have\footnote{%
  Markov inequality implies for at least $\frac23$ of the
  codes in the ensemble,
  \begin{equation*}
    \frac{1}{2M} \sum_{w=1}^{2M} \Pr[\hat{W} \ne w | W = w] \le 3 \EE\Bigl[
    \frac{1}{2M} \sum_{w=1}^{2M} \Pr[\hat{W} \ne w | W = w] \Bigr].
  \end{equation*}
  Similarly for at least $\frac23$ of the codes in the ensemble,
  \begin{equation*}
    \frac{1}{2M} \sum_{w=1}^{2M} \KLDiv(P_{\bZ|W=w} \Vert P_Z^n) \le 3 \EE\Bigl[
    \frac{1}{2M} \sum_{w=1}^{2M} \KLDiv(P_{\bZ|W=w} \Vert P_Z^n) \Bigr].
  \end{equation*}
  Therefore, for at least $\frac13$ of the codes in the ensemble both
  \eqref{eq:averagePe} and \eqref{eq:averageD} hold simultaneously.
}:
\begin{align}
  & \frac{1}{2M} \sum_{w=1}^{2M} \Pr[\hat{W} \ne w | W = w]  \nonumber \\
  & \qquad \dotle \exp[-n E_{\rm r}(P_X,\sfV, R+R')].
  \label{eq:averagePe} \\
  & \frac1{2M}\sum_{w=1}^{2M} \KLDiv(P_{\bZ|W=w} \Vert
  P_Z^n) \nonumber \\
  & \qquad \dotle \exp[-n E_{\rm s}(P_X,\sfW,R')]. \label{eq:averageD}
\end{align}

Since each of the summands in \eqref{eq:averagePe} is positive, there exist a
subset $\calW_1  \subset \{1,\dots,2M\}$ of cardinality $\abs{\calW_1} \ge
\frac32 M$ such that, for $\forall w \in \calW_1$,
\begin{equation}
  \Pr[\hat{W} \ne w | W = w] \dotle 4 \exp[-n E_{\rm r}(P_X,\sfV,R+R')].
  \label{eq:maxPe}
\end{equation}
Similarly since the summands in \eqref{eq:averageD}
are positive, there exists a subset $\calW_2 \subset \{1,\dots,2M\}$ of
cardinality $\abs{\calW_2} \ge \frac32 M$ such that, for $\forall w \in
\calW_2$, 
\begin{equation}
  \KLDiv(P_{\bZ|W=w} \Vert P_Z^n) \dotle 4 \exp[-n E_{\rm s}(P_X,\sfW,R')].
  \label{eq:maxD}
\end{equation}

Since $\abs{\calW_1 \cap \calW_2} \ge M$ there exist a
subset $\calW \subseteq \calW_1 \cap \calW_2$ of cardinality $\abs{\calW} = M$. 
The sub-code defined by the messages in $\calW$, $\bigcup_{w \in \calW}
\calC_w^*$ has rate $R$ and, using that, for any message distribution $P_W$ on
$\calW$, we have:
\begin{align*}
  P_{\rm e} & = \sum_{w \in \calW} P_W(w) \Pr[\hat{W} \ne w | W = w] \nonumber
  \\
  & \dotle \exp[-n E_{\rm r}(P_X,\sfV,R+R')],
\end{align*}
due to \eqref{eq:maxPe}, and
\begin{align*}
  I(W;\bZ) & = \KLDiv(P_{\bZ|W} \Vert P_Z^n | P_W) - \KLDiv(P_{\bZ} \Vert
  P_Z^n) \\ 
  & \le \sum_{w \in \calW} P_W(w) \KLDiv\bigl(P_{\bZ|W=w} \Vert P_Z^n \bigr) \\
  & \dotle \exp[-n E_s(R',P_X,\sfW)],
\end{align*} 
due to \eqref{eq:maxD}. \hfill \IEEEQED
\section{Derivation of Exponents for The Proof of Lemma~\ref{lem:typeBound}}
\subsection{Derivation of $E_b$ and It's Properties} \label{app:eb}
\begin{proposition} Let $E_b(P_{X,Z}, P, a)$ be defined as in \eqref{eq:EbDef}.
  Then,
  \begin{equation}
    E_b(P_{X,Z}, P, a) = a + \max_{\rho \in \RRR} \bigl\{\rho a -
      G_0(P_{X,Z},P,\rho) \bigr\}, \label{eq:EbVal}
  \end{equation}
  where $G_0$ is defined in \eqref{eq:g0def}. 
\end{proposition}
\begin{IEEEproof} 
  Let 
  \begin{equation*}
    \iota_{X,Z} (x,z) \triangleq \log\left( \frac{P_{X,Z}(x,z)}{P_X(x) P_Z(z)}
      \right), \qquad \forall (x,z) \in \calX \times \calZ,
  \end{equation*}
  denote the information density function for the joint distribution $P_{X,Z}$
  for the sake of brevity.

  Using \eqref{eq:apqdef},
  \begin{align}
    & \min_{\hat\sfQ: A_{X,Z}(P;\hat\sfQ)=a} \KLDiv(\hat\sfQ \Vert P_X |
    P)
    \nonumber \\ 
    & \qquad \qquad =  a +  \min_{\hat\sfQ: A_{X,Z}(P;\hat\sfQ)=a}
    \KLDiv(\hat\sfQ \Vert P_{X|Z} | P)
    \label{eq:ebsum}
  \end{align}
  Now,  we have
  \begin{align*}
    & \min_{\hat\sfQ: A_{X,Z}(P;\hat\sfQ)=a} \KLDiv(\hat\sfQ \Vert P_{X|Z}
    |
    P)
    \\ 
    & \qquad = \min_{\hat\sfQ} \Bigl\{
      \KLDiv(\hat\sfQ \Vert P_{X|Z} | P) + \max_{\rho \in \RRR} \rho
      \bigl( a - A_{X,Z}(P; \hat\sfQ) \bigr)
    \Bigr\} 
    \\ 
    & \qquad = \min_{\hat\sfQ} \max_{\rho \in \RRR} \Bigl\{
      \KLDiv(\hat\sfQ \Vert P_{X|Z} | P) + \rho
      \bigl( a - A_{X,Z}(P; \hat\sfQ) \bigr)
    \Bigr\} 
    \\ 
    & \qquad \stackrel{(*)}{=} \max_{\rho \in \RRR} \Bigl\{ \min_{\hat\sfQ}
    \bigl\{
      \KLDiv(\hat\sfQ \Vert P_{X|Z} | P) - \rho A_{X,Z}(P; \hat\sfQ)
    \bigr\}
    + \rho a \Bigr\} 
  \end{align*}
  where (*) follows since $\KLDiv(\hat\sfQ \Vert P_{X|Z} | P)$ is a convex
  function of $\hat\sfQ$ and $A_{X,Z}(P;\hat\sfQ)$ is a linear function of
  $\hat\sfQ$. Therefore, $\KLDiv(\hat\sfQ \Vert P_{X|Z} | P) - \rho
  A_{X,Z}(P;\hat\sfQ)$ is also a convex function of $\hat\sfQ$ and we can
  swap the $\min$ and the $\max$. Now,
  \begin{align*}
    & \KLDiv(\hat\sfQ \Vert P_{X|Z} | P) - \rho A_{X,Z}(\hat P; \hat\sfQ)
    \\
    & \quad = \sum_{z \in \calZ} P(z) \sum_{x \in \calX} \hat\sfQ(x|z)
    \log \Bigl( \frac{\hat\sfQ(x|z)}{P_{X|Z}(x|z) \exp[\rho \iota_{X,Z}(x,z)]}
    \Bigr) \\
    & \quad \ge \sum_{z \in \calZ} P(z)  \log \biggl( \frac{1}{
    \sum_{x \in \calX} P_{X|Z}(x|z) \exp[\rho \iota_{X,Z}(x,z)]} \biggr)
  \end{align*}
  with equality iff $\hat\sfQ(x|z) \propto P_{X|Z}(x|z) \exp[\rho
  \iota_{X,Z}(x,z)]$ (using the concavity of logarithm). Therefore,
  \begin{align}
    & \min_{\hat\sfQ} \bigl\{ \KLDiv(\hat\sfQ \Vert P_{X|Z} | P) - \rho
    A_{X,Z}(P;\hat\sfQ) \bigr\} + \rho a = \rho a  \nonumber \\
    & \quad  - \sum_{z \in \calZ} P(z) \log \Bigl( \sum_{x
    \in \calX} P_{X|Z}(x|z) \exp[\rho \iota_{X,Z}(x,z)] \Bigr).
    \nonumber \tag*{\IEEEQEDhere}
  \end{align}
\end{IEEEproof}
\begin{remark}
  It is easy to verify that $E_b(P_{X,Z}, P,a)$
  is a convex function of $a$. Furthermore, \eqref{eq:ebsum} implies
  $E_b(P_{X,Z},P,a) \ge a$ with equality at $a = \KLDiv(P_{X|Z} \Vert
  P_X | P)$. 
\end{remark}
\subsection{Derivation of $E_1$ and $E_2$} \label{app:e1e2}
\begin{IEEEproof}[Proof of \eqref{eq:e1val}]
  Using \eqref{eq:e1def},
  \begin{align}
    & E_1(P_{X,Z},R', P) = \min_{a \le R'} \bigl\{ R' + E_b(P_{X,Z}, P, a) - 2 a
    \bigr\} \nonumber \\
    & \, = \min_{a \in \RRR} \Bigl\{ R' + E_b(P_{X,Z},P, a) - 2 a +
    \max_{\lambda \le 0} \lambda(R' - a) \Bigr\} \nonumber \\
    & \, = \min_{a \in \RRR} \max_{\lambda \le 0} \Bigl\{ (1+\lambda) R' +
    E_b(P_{X,Z}, P, a) - (2+\lambda) a \Bigr\} \nonumber \\
    & \quad = \min_{a \in \RRR} \max_{\lambda \le 1} \Bigl\{ \lambda R' +
    E_b(P_{X,Z}, P, a) - (1+\lambda) a \Bigr\} \nonumber \\
    & \, \stackrel{(*)}{=} \max_{\lambda \le 1} \Bigl\{ \lambda R' + \min_{a
    \in \RRR} \bigl\{E_b(P_{X,Z},P,a) - (1+\lambda) a \bigr\} \Bigl\},
    \label{eq:e1maxmin1}
  \end{align}
  where (*) follows since $E_b(P_{X,Z}, P, a)$ is convex in $a$.  Using
  \eqref{eq:EbVal} we have
  \begin{align} 
    & \min_{a \in \RRR} \bigl\{E_b(P_{X,Z},P,a) - (1+\lambda) a \bigr\}
    \nonumber \\ 
    & \quad = \min_{a \in \RRR} \Bigl\{ \max_{\rho \in \RRR} \bigl\{ \rho a -
    G_0(P_{X,Z}, P, \rho) \bigr\} - \lambda a \Bigl\} \nonumber \\
    & \quad = \min_{a\in \RRR} \max_{\rho \in \RRR} \bigl\{ (\rho - \lambda) a -
    G_0(P_{X,Z},P, \rho) \bigr\} \nonumber \\
    & \quad \stackrel{(*)}{=} \max_{\rho \in \RRR} \Bigl\{ \min_{a \in \RRR}
    \bigl\{(\rho - \lambda) a \bigr\} - G_0(P_{X,Z},P,\rho) \Bigr\},
    \label{eq:e1maxmin2}
  \end{align}
  where again (*) follows since $G_0(P_{X,Z}, P, \rho)$ is convex in $\rho$ (cf.
  Appendix~\ref{app:g0convex}). We
  then note that the minimum of the linear term $(\rho - \lambda) a$ over the
  choices of $a$ is $-\infty$ unless $\rho = \lambda$. Therefore, the result of
  \eqref{eq:e1maxmin2} is 
  \begin{equation}
    \min_{a \in \RRR} \bigl\{E_b(P_{X,Z},P,a) - (1+\lambda) a \bigr\} = -
    G_0(P_{X,Z}, P, \lambda) \label{eq:e1maxmin2val}
  \end{equation}
  Plugging the above into \eqref{eq:e1maxmin1} completes the proof.
\end{IEEEproof}
\begin{IEEEproof}[Proof of \eqref{eq:e2val}]
  Similarly, using \eqref{eq:e2def},
  \begin{align}
    & E_2(P_{X,Z},R', P) = \min_{a > R'} \bigl\{ E_b(P_{X,Z}, P, a) -  a
    \bigr\} \nonumber \\
    & \, = \min_{a \in \RRR} \Bigl\{ E_b(P_{X,Z},P, a) -  a +
    \max_{\lambda \ge 0} \lambda(R' - a) \Bigr\} \nonumber \\
    & \, = \min_{a \in \RRR} \max_{\lambda \ge 0} \Bigl\{ \lambda R' +
    E_b(P_{X,Z}, P, a) - (1+\lambda) a \Bigr\} \nonumber \\
    & \, \stackrel{(*)}{=} \max_{\lambda \ge 0} \Bigl\{ \lambda R' + \min_{a
    \in \RRR} \bigl\{E_b(P_{X,Z},P,a) - (1+\lambda) a \bigr\} \Bigl\},
    \label{eq:e2maxmin}
  \end{align}
  where (*) follows since $E_b(P_{X,Z}, P, a)$ is convex in $a$.  Using
  \eqref{eq:e1maxmin2val} in \eqref{eq:e2maxmin} completes the proof.
\end{IEEEproof}
\section{Derivation of $E_{\rm s}$} \label{app:secExp}
Plugging \eqref{eq:EtVal} into \eqref{eq:exponentDef} we have
\begin{align}
  & \min_{P \in \calP(\calZ)} \bigl\{ E_t(P_{X,Z}, P, R') +
  \KLDiv(P \Vert P_Z ) \bigr\}  \nonumber \\
  & \quad =  
  \min_{P \in \calP(\calZ)} \bigl\{
    \max_{0 \le \lambda \le 1} \{ \lambda R' -
      G_0(P_{X,Z}, P, \lambda) 
    \} +
    \KLDiv(P \Vert P_Z )
  \bigr\} \nonumber \\
  & \quad \stackrel{(*)}{=} 
  \max_{0 \le \lambda \le 1} \bigl\{ \lambda R' 
    +
    \min_{P \in \calP(\calZ)} \{ \KLDiv(P \Vert P_Z) -
      G_0(P_{X,Z},P, \lambda) 
    \}
  \bigr\} \nonumber
\end{align}
where (*) follows since $G_0$, defined in \eqref{eq:g0def} is a linear function
of $P$ while $\KLDiv(\hat P \Vert P_Z)$ is convex in $P$ and we can swap the
$\min$ and the $\max$. The claim follows then by observing that
\begin{align}
  & \KLDiv(P\Vert P_Z) - G_0(P_{X,Z}, P, \lambda) \nonumber \\
  &  \, = \sum_{z \in \calZ} P(z) \Biggl[%
    \log \left( \frac{P(z)}{P_Z(z)} \right) - 
    \nonumber \\
    & \quad \qquad -
    \log \Bigl( 
    \sum_{x \in\calX} P_{X|Z}(x|z)^{1+\lambda} P_X(x|z)^{-\lambda}
    \Bigr)
  \Biggr] \nonumber \\ 
  & \, \ge 
  \log \left[ \frac{1}{\displaystyle \sum_{z \in \calZ} P_Z(z) \sum_{x \in
  \calX} \Bigl( P_{X|Z}(x|z)^{1+\lambda}
  P_X(x)^{-\lambda} \Bigr)}
  \right], \nonumber 
\end{align}
with equality if 
\begin{equation*}
  P(z) \propto  P_Z(z) \sum_{x \in \calX} \Bigl(
  P_{X|Z}(x|z)^{1+\lambda}
  P_X(x)^{-\lambda} \Bigr),
\end{equation*} 
using the concavity of logarithm. \hfill \IEEEQEDhere
\section{Convexity Proofs}
\begin{lemma} \label{lem:holder}
  Let $a_i > 0,$  and $b_i \ge 0, i = 1,\dots,k$ be arbitrary real numbers.
  Then the function 
  \begin{equation*}
    f(s) \triangleq \log\Bigl( \sum_{i=1}^k a_i b_i^s \Bigr),
  \end{equation*}
  is convex in $s$ for $\forall s \in \RRR$. 
\end{lemma}
\begin{IEEEproof}
  Pick $s_1 < s_2$ and $t \in (0,1)$. Let $\bart \triangleq 1-t$ and $s
  \triangleq t s_1 + \bart s_2$. Then, H\"older's inequality implies
  \begin{equation*}
    \sum_{i=1}^{k} a_i b_i^s  = \sum_{i=1}^k \Bigl(
    a_i^t b_i^{t s_1} \times a_i^{\bart} b_i^{\bart s_1} \Bigr) 
     \le 
    \Bigl(\sum_{i=1}^k a_i b_i^{s_1}\Bigr)^t \Bigl(\sum_{i=1}^{k} a_i
    b_i^{s_2}\Bigr)^{\bart}.
  \end{equation*}
  Taking the $\log$ of both sides of the above concludes the proof.
\end{IEEEproof}
\begin{lemma} \label{lem:convexSum}
  Suppose $f_i(s), i = 1,2,\dots,k$ are convex functions in $s$
  and $a_i > 0, i = 1,2, \dots, k$ is a sequence of real numbers. Then,
  \begin{enumerate}[(i)]
    \item $f(s) \triangleq \sum_{i=1}^k a_i f_i(s)$ is convex in $s$.
    \item $g(s) \triangleq \log \Bigl( \sum_{i=1}^k a_i \exp[f_i(s)] \Bigr)$ is
      convex in $s$.
  \end{enumerate}
\end{lemma}
\begin{IEEEproof}
  The convexity of $f(s)$ is trivial. To prove the convexity of $g(s)$, let $s_1
  < s_2$ and $s = t s_1 + \bart s_2$ for some $t \in (0,1)$ (where $\bart
  \triangleq 1-t$). Then
  \begin{align*}
    & \sum_{i=1}^k a_i \exp[f_i(s)] \le \sum_{i=1}^k a_i \exp[t f_i(s_1) +
    (1-t) f_i(s_2)] \\
    & \quad = 
    \sum_{i=1}^k \Bigl( a_i^t \exp[t f_i(s_1)] \times  a_i^{\bart} \exp[\bart
    f_i(s_2)] \Bigr) \\
    & \quad \le \Bigl(\sum_{i=1}^k a_i \exp[f_i(s_1)] \Bigr)^t
    \Bigl(\sum_{i=1}^k a_i \exp[f_i(s_2)]\Bigr)^{\bart}
  \end{align*}
  where the second inequality follows by H\"older's inequality. Taking the
  logarithm of both sides of the above proves (ii).
\end{IEEEproof}
Convexity of the functions $F_0$ and $G_0$ is established using the above two
lemmas as follows:
\subsection{Convexity of $G_0$} \label{app:g0convex}
Set $a_i = P_{X|Z}(x|z)$ and $b_i =
\frac{P_{X|Z}(x|z)}{P_X(x)}$ in Lemma~\ref{lem:holder} and then use
Lemma~\ref{lem:convexSum} part (i). \hfill\IEEEQED
\subsection{Convexity of $F_0$} \label{app:f0convex}
Set $a_i = P_{X|Z}(x|z)$ and $b_i = \frac{P_{X|Z}(x|z)}{P_X(x)}$ in
Lemma~\ref{lem:holder} and then use Lemma~\ref{lem:convexSum} part
(ii). \hfill\IEEEQED\\
\end{document}